\documentclass[aps,prd,preprint,grouptedaddress,nofootinbib,showpacs,showkeys]{revtex4}
\usepackage{graphicx}
\begin{document}

\preprint{UTAP-453}
\preprint{hep-ph/0307169}

\title{Decaying neutrinos and implications from the supernova relic
neutrino observation}


\author{Shin'ichiro Ando}
\email[Email address: ]{ando@utap.phys.s.u-tokyo.ac.jp}
\affiliation{Department of Physics, School of Science, The University of
Tokyo, 7-3-1 Hongo, Bunkyo-ku, Tokyo 113-0033, Japan}


\date{Received 27 May 2003; received in revised form 5 July 2003;
accepted 10 July 2003}

\begin{abstract}

We propose that supernova relic neutrino (SRN) observation can be used
 to set constraints on the neutrino decay models.
Because of the long distance scale from cosmological supernovae to the
 Earth, SRN have possibility to provide much stronger limit than the
 present one obtained from solar neutrino observation.
Since the currently available data are only the upper limit on the flux
 integrated over $E_{\bar\nu_e}>19.3$ MeV, the decay models on which we
 can set constraints is quite restricted; they must satisfy specific
 conditions such that the daughter neutrinos are active species, the
 neutrino mass spectrum is quasi-degenerate, and the neutrino mass
 hierarchy is normal.
Our numerical calculation clearly indicates that the neutrino decay
 model with $(\tau_2/m,\tau_3/m)\alt (10^{10},10^{10})$ [s/eV], where
 $\tau_i$  represents the lifetime of mass eigenstates $\bar\nu_i$,
 appears to give the SRN flux that is larger than the current upper
 limit.
However, since the theoretical SRN prediction contains many
 uncertainties concerning a supernova rate in the universe or simulation
 of supernova explosions, we cannot conclude that there exists the
 excluded parameter region of the neutrino lifetime.
In the near future, further reduced upper limit is actually expected,
 and it will provide more severe constraints on the neutrino decay
 models.
\end{abstract}


\maketitle

\section{Introduction\label{sec:Introduction}}

A number of ground-based experiments, which observed atmospheric
\citep{Fukuda99}, solar \citep{Fukuda02a,Ahmad02a,Ahmad02b}, and reactor
neutrinos \citep{Eguchi03}, have revealed the nonzero neutrino masses
and flavor mixings, i.e., properties beyond the standard model of the
particle physics.
Fortunately, our current knowledge of the neutrino mass differences and
mixing angles further enables us to consider far more exotic properties
of the neutrino, such as a nonzero magnetic moment (see
Refs. \citep{Ando03b,Ando03d} and references therein) and neutrino
decay.
In this Letter we show that neutrino decay of a particular type may be
ruled out or severely constrained by the current and the future
supernova relic neutrino observation at the Super-Kamiokande (SK)
detector \citep{Malek03}.

We consider two-body neutrino decays such as $\nu_i\to\nu_j+X$, where
$\nu_i$ are neutrino mass eigenstates and $X$ denotes a very light or
massless particle, e.g., a Majoron.
The strongest limits on this decay modes are obtained from the solar
neutrino observation by SK \citep{Beacom02} (see also
Ref. \citep{Bandyopadhyay03}).
It was argued that the limit is obtained primarily by the nondistortion
of the solar neutrino spectrum, and the potentially competing
distortions caused by oscillations as well as the appearance of active
daughter neutrinos were also taken into account.
However, owing to the restricted distance scale to the Sun, this limit
is very weak, $\tau/m\agt 10^{-4}~{\rm s/eV}$, and therefore, the
possibility of the other astrophysical neutrino decay via the same modes
cannot be eliminated.
In fact, detectability of the decay of neutrinos from the high-energy
astrophysical sources was discussed \citep{Beacom03}, and it has been
concluded that it should be visible by future km-scale detectors such as
IceCube, since the neutrino decay strongly alter the flavor ratios from
the standard one, $\phi_{\nu_e}:\phi_{\nu_\mu}:\phi_{\nu_\tau}=1:1:1$,
expected from oscillations alone.

Our strategy is basically the same as that of the previous studies
\citep{Beacom02,Beacom03}, i.e., the enhancement of the electron
neutrino events due to the decay is investigated.
As a source of neutrinos, we consider supernovae.
Two observational results concerning supernova neutrinos exist; one is
the well-known neutrino burst from SN 1987A \citep{Hirata87,Bionta87},
and the other is the recent upper limit on the flux of supernova relic
neutrinos (SRN), which is the accumulation of neutrinos from all the
past supernovae, by SK \citep{Malek03}.\footnote{From this point on, we
consider only $\bar\nu_e$ at detection because this kind of flavor is
most efficiently detected at SK via $\bar\nu_ep\to e^+n$ reaction.}
Original discussions concerning the SN 1987A signal have been already
given, including the effect of the neutrino decay as well as the pure
flavor mixing, in the literatures \citep{Frieman88,Raghavan88} (see also
Ref. \citep{Lindner02}).
In this Letter, we use the latter one (SRN) for obtaining implications
for the neutrino decay.
The advantage of this approach compared to the former ones is that the
neutrinos must transit over very long distance scale from the
cosmological supernovae to the Earth, and much longer lifetimes would be
probed {\it in principle}.
The current SRN upper limit is only a factor three larger than
theoretical predictions by Ando et al. \citep{Ando03a,Ando03c}
(hereafter AST), which adopted neutrino oscillations using
experimentally inferred parameters.
Therefore, if some neutrino decay model predicts the SRN flux which is
three times larger than the AST prediction, then it is excluded.

This Letter is organized as follows.
In Section \ref{sec:Formulation and models}, we present formulation for
the SRN calculation and adopted models.
In particular the detailed discussion concerning the supernova model
and the neutrino decay model are given in Section \ref{sub:Supernova
model} and \ref{sub:Neutrino decay model}, respectively.
How the oscillation and decay changes the neutrino spectrum and flux is
qualitatively illustrated in Section \ref{sec:Neutrino oscillation and
decay} and the numerically calculated results are given in Section
\ref{sec:Results}.
In Section \ref{sec:Discussion}, we discuss the various possibilities of
the neutrino decay, uncertainties of the adopted models, and future
prospects.

\section{Formulation and models \label{sec:Formulation and models}}

The SRN $\bar\nu_e$ flux is calculated by
\begin{equation}
 \frac{dF_{\bar\nu_e}}{dE_{\bar\nu_e}}
  =c\int_0^{z_{\rm max}}R_{\rm SN}(z)\mathcal N_{\bar\nu_e}
  (E_{\bar\nu_e}^\prime,z)(1+z)\frac{dt}{dz}dz,
  \label{eq:SRN flux}
\end{equation}
where $E_\nu^\prime=(1+z)E_\nu$, $R_{\rm SN}(z)$ is a supernova rate
per comoving volume at redshift $z$, and $z_{\rm max}$ is the redshift
when the gravitational collapses began (we assume it to be 5).
$\mathcal N_\nu(E_\nu^\prime,z)=dN_\nu(E_\nu^\prime)/dE_\nu^\prime$ is
the number of emitted neutrinos per unit energy range by one supernova
at redshift $z$.
As the supernova rate, we use the most reasonable model to date, which
is based on the rest-frame UV observation of star formation history in
the universe by the Hubble Space Telescope \cite{Madau96}, and the model
was also used in AST as ``SN1.''
In this model, the supernova rate exponentially increases with $z$,
peaks around $z\sim 1.5$, and exponentially decreases in further
high-$z$ region.

\subsection{Supernova model \label{sub:Supernova model}}

As original neutrino spectra $\mathcal N_{\nu}^0(E,z)$, which is not the
same as $\mathcal N_{\nu}$ owing to the neutrino oscillation and decay,
we adopt the result of a numerical simulation by the Lawrence Livermore
group \citep{Totani98}, which is the only group that is successful for
calculating neutrino luminosities during the entire burst ($\sim 10$
sec).
The average energies are different between flavors, such as $\langle
E_{\nu_e}\rangle\simeq 11~{\rm MeV},\langle E_{\bar\nu_e}\rangle\simeq
16~{\rm MeV},\langle E_{\nu_x}\rangle\simeq 22~{\rm MeV}$, where $\nu_x$
represents non-electron neutrinos and antineutrinos.
This hierarchy of the average energies is explained as follows.
Since $\nu_x$ interact with matter only through the neutral-current
reactions in supernova, they are weakly coupled with matter compared to
$\nu_e$ and $\bar\nu_e$.
Thus the neutrino sphere of $\nu_x$ is deeper in the core than that of
$\nu_e$ and $\bar\nu_e$, which leads to higher temperatures for
$\nu_x$.
The difference between $\nu_e$ and $\bar\nu_e$ comes from the fact that
the core is neutron-rich and $\nu_e$ couple with matter more strongly,
through $\nu_en\leftrightarrow e^-p$ reaction.

Our calculations presented in Section \ref{sec:Results} are strongly
sensitive to the adopted supernova model, or in particular, the average
energy difference between $\bar\nu_e$ and $\nu_x$.
Recently, the Livermore calculation is criticized since it lacks the
relevant neutrino processes such as neutrino bremsstrahlung and
neutrino-nucleon scattering with nucleon recoils, which are considered
to make the mean energy difference between flavors less prominent; it
has been actually confirmed by the recent simulations (e.g.,
Ref. \citep{Thompson02}).
However, we cannot adopt recent models, even though they include all the
relevant neutrino microphysics.
This is because the SRN calculation definitely requires the
time-integrated neutrino spectrum during the entire burst, whereas all
the recent supernova simulations terminate at $\sim 0.5$ sec after the
core bounce.
Since the Livermore group alone is successful to simulate the supernova
explosion and calculate the neutrino luminosity during the entire burst,
we use their result as a reference model.
Again, we should note that our discussions from this point on heavily
relies on the adopted supernova model.

\subsection{Neutrino decay model \label{sub:Neutrino decay model}}

In this Letter, we consider only the so-called ``invisible'' decays,
i.e., decays into possibly detectable neutrinos plus truly invisible
particles, e.g., light scalar or pseudoscalar bosons.
The best limit on the lifetime with this mode is obtained from solar
neutrino observations and is $\tau/m\agt 10^{-4}$ s/eV
\citep{Beacom02}, which is too small to set relevant constraints on
discussions below.
We do not consider other modes such as radiative two-body decay since
they are experimentally constrained to have very long life times (e.g.,
Ref. \citep{Groom00}).

We note that our approach is powerful only when the decay model
satisfies specific conditions such that (i) the daughter neutrinos are
active species, (ii) the neutrino mass spectrum is quasi-degenerate
($m_1\approx m_2\approx m_3$), and (iii) the neutrino mass hierarchy is
normal ($m_1<m_3$), not inverted ($m_1>m_3$).
This is because at present, only the upper limit of the SRN flux is
obtained, and therefore, the decay model which does not give large flux
at detection energy range ($E_{\bar\nu_e}>19.3$ MeV) cannot be
satisfactorily constrained.
All these conditions (i)--(iii) must be satisfied to obtain severe
constraints on the neutrino lifetime, because
(i) if the daughter neutrinos are sterile species, the SRN flux
decreases compared with the model without the neutrino decay;
(ii) if the neutrino mass spectrum is strongly hierarchical, then the
daughter neutrino energy is degraded compared to its parent and the
predicted SRN flux at high energy range would not be large;
and (iii) in the case of inverted hierarchy, $\bar\nu_e$ most tightly
couple to the heaviest mass eigenstates, which decay into lighter
states, and it also reduces the SRN flux.

For a while, we assume that the conditions (i)--(iii) are satisfied;
all the other possibilities are addressed in detail in Section
\ref{sub:Other decaying modes}.
As discussed in Section \ref{sub:Future prospects}, we believe that
future observational development would enable far more general and
model-independent discussions, which are not restricted by the above
conditions.

\section{Neutrino oscillation and decay \label{sec:Neutrino oscillation
and decay}}

Effects of neutrino oscillation and decay are included in $\mathcal
N_{\bar\nu_e}(E,z)$.
We address the problem of flavor mixing by the pure supernova matter
effect, first for antineutrinos and second for neutrinos, and then we
discuss the neutrino decay.

The state of $\bar\nu_e$ produced at deep in the core is coincident with
the lightest mass eigenstate $\bar\nu_1$, owing to large matter
potential.
This state propagates to the supernova surface without being influenced
by level crossings between different mass eigenstates (it is said that
there are no resonance points).
Thus, $\bar\nu_e$ at production becomes $\bar\nu_1$ at the stellar
surface $(\mathcal N_{\bar\nu_1}=\mathcal N_{\bar\nu_e}^0)$ and the number
of $\bar\nu_e$ there is given by
\begin{eqnarray}
 \mathcal N_{\bar\nu_e}(E,z)
 &\simeq&\cos^2\theta_\odot\mathcal N_{\bar\nu_1}(E,z)
  +\sin^2\theta_\odot\mathcal N_{\bar\nu_2}(E,z)\nonumber\\
 &=&\cos^2\theta_\odot\mathcal N_{\bar\nu_e}^0(E,z)
  +\sin^2\theta_\odot\mathcal N_{\nu_x}^0(E,z),\nonumber\\
 \label{eq:pure oscillation}
\end{eqnarray}
where $\theta_\odot$ is the mixing angle inferred from the solar
neutrino observations ($\cos^2\theta_\odot\simeq 0.7$)
\citep{Fukuda02a,Ahmad02a,Ahmad02b,Eguchi03}, $\mathcal N^0_{\nu}$
represents the neutrino spectrum at production.
In the above expression (\ref{eq:pure oscillation}), we used the fact,
$\sin^22\theta_{\rm atm}=1$, and assumed $\theta_{13}=0$.
These are justified by the atmospheric \citep{Fukuda99} and reactor
neutrino experiments \citep{Apollonio99}.

The situations changes dramatically for neutrino sector.
As the case for antineutrinos, $\nu_e$ are produced as mass eigenstates
owing to large matter potential, however, the difference is that the
produced $\nu_e$ coincide with the heaviest state $\nu_3$.
Since in vacuum $\nu_e$ most strongly couples to the lightest state
$\nu_1$, there must be two level crossings (or resonance points) between
different mass eigenstates during the propagation through supernova
envelope; each is labeled by H- and L-resonance, corresponding to
whether the density of the resonance point is higher or lower (see,
e.g., Ref. \citep{Dighe00} for details).
It is well-known that at L-resonance the mass eigenstate does not flip
(adiabatic resonance) for LMA solution to the solar neutrino problem.
However, the adiabaticity of the H-resonance becomes larger than unity
when the parameter $\theta_{13}$ is sufficiently large.
Instead, we parameterize the flip probability at the H-resonance by
$P_H$, i.e., if the resonance is adiabatic (nonadiabatic), $P_H=0(1)$.
Thus, at the stellar surface, the neutrino spectra of mass eigenstates
is given by
\begin{eqnarray}
 \mathcal N_{\nu_1}(E,z)&=&\mathcal N_{\nu_x}^0(E,z),
 \label{eq:nu_1}\\
 \mathcal N_{\nu_2}(E,z)&=&P_H\mathcal N_{\nu_e}^0(E,z)
  +(1-P_H)\mathcal N_{\nu_x}^0(E,z),
  \label{eq:nu_2}\\
 \mathcal N_{\nu_3}(E,z)&=&(1-P_H)\mathcal N_{\nu_e}^0(E,z)
  +P_H\mathcal N_{\nu_x}^0(E,z).
  \label{eq:nu_3}
\end{eqnarray}

If we include the neutrino decay, the expected $\bar\nu_e$ flux from
each supernova changes drastically.
Before giving a detailed discussion, we first place some simplifying
assumptions.
Instead of the lifetime, we define ``decay redshift'' $z_i^d$ of
the mass eigenstate $\nu_i(\bar\nu_i)$; if the source redshift $z$ is
larger than the decay redshift $z_i^d$, all the neutrinos $\nu_i
(\bar\nu_i)$ decay, on the other hand if $z<z_i^d$, $\nu_i(\bar\nu_i)$
completely survive.
We consider the decaying mode $\nu_3(\bar\nu_3)\to\bar\nu_1$ and $\nu_2
(\bar\nu_2)\to\bar\nu_1$, and $z_2^d$ and $z_3^d$ are taken to be two
free parameters.
The other case that one of them is stable can be realized if we take
$z^d>z_{\rm max}$.
With these assumptions and parameterization, the neutrino spectrum which
is emitted by the source at redshift $z$ can be obtained.
First, we consider the decay mode $\bar\nu_i\to\bar\nu_j +X$, in which
the neutrino helicity conserves.
The $\bar\nu_e$ spectrum is given by
\begin{eqnarray}
 \mathcal N_{\bar\nu_e}(E,z)&=&
  \cos^2\theta_\odot\mathcal N_{\bar\nu_1}(E,z)
  +\sin^2\theta_\odot\mathcal N_{\bar\nu_2}(E,z)\nonumber\\
 &=&\cos^2\theta_\odot
  \left[\mathcal N_{\bar\nu_1}^0(E,z)
   +\mathcal N_{\bar\nu_2}^0(E,z)\Theta(z-z_2^d)\right.\nonumber\\
 &&{}\left.+\mathcal N_{\bar\nu_3}^0(E,z)\Theta(z-z_3^d)\right]
\nonumber\\
 &&{}+\sin^2\theta_\odot\mathcal N_{\bar\nu_2}^0(E,z)
  \Theta(z_2^d-z)\nonumber\\
 &=&\cos^2\theta_\odot
  \left[\mathcal N_{\bar\nu_e}^0(E,z)
   +\mathcal N_{\nu_x}^0(E,z)\right.\nonumber\\
 &&{}\left.\times\left\{\Theta(z-z_2^d)
	    +\Theta(z-z_3^d)\right\}\right]
	    \nonumber\\
 &&{}+\sin^2\theta_\odot\mathcal N_{\nu_x}^0(E,z)
  \Theta(z_2^d-z),
  \label{eq:decayed spectrum}
\end{eqnarray}
where $\Theta$ is the step function.
On the other hand, if the relevant decay mode is $\nu_i(\bar\nu_i)\to
\bar\nu_j(\nu_j) +X$ (helicity flips), the expected spectrum becomes
\begin{eqnarray}
 \mathcal N_{\bar\nu_e}(E,z)&=&
  \cos^2\theta_\odot\mathcal N_{\bar\nu_1}(E,z)
  +\sin^2\theta_\odot\mathcal N_{\bar\nu_2}(E,z)\nonumber\\
 &=&\cos^2\theta_\odot
  \left[\mathcal N_{\bar\nu_1}^0(E,z)
   +\mathcal N_{\nu_2}^0(E,z)\Theta(z-z_2^d)\right.\nonumber\\
 &&{}\left.+\mathcal N_{\nu_3}^0(E,z)\Theta(z-z_3^d)\right]
\nonumber\\
 &&{}+\sin^2\theta_\odot\mathcal N_{\bar\nu_2}^0(E,z)
  \Theta(z_2^d-z)\nonumber\\
 &=&\cos^2\theta_\odot
  \left[\mathcal N_{\bar\nu_e}^0(E,z)
   +\mathcal N_{\nu_e}^0(E,z)\right.\nonumber\\
 &&{}\times\left\{P_H\Theta(z-z_2^d)
	    +(1-P_H)\Theta(z-z_3^d)\right\}\nonumber\\
 &&{}+\mathcal N_{\nu_x}^0(E,z)
  \left\{(1-P_H)\Theta(z-z_2^d)\right.\nonumber\\
 &&{}\left.\left.+P_H\Theta(z-z_3^d)\right\}\right]
	    \nonumber\\
 &&{}+\sin^2\theta_\odot\mathcal N_{\nu_x}^0(E,z)
  \Theta(z_2^d-z).
  \label{eq:decayed spectrum from neutrinos}
\end{eqnarray}
Comparing Eqs. (\ref{eq:decayed spectrum}) and (\ref{eq:decayed spectrum
from neutrinos}) with Eq. (\ref{eq:pure oscillation}), the SRN flux with
the neutrino decay is expected to be very different from the case of the
pure neutrino oscillation.
In particular, for the mode $\bar\nu_i\to\bar\nu_j +X$
[Eq. (\ref{eq:decayed spectrum})], the SRN spectrum is expected to be
hard since it contains a fair amount of $\nu_x$.
In the case of $\nu_i\to\bar\nu_j +X$ mode, the $\nu_e$ spectrum is also
included, and then the corresponding upper limit is not as strong as
that for $\bar\nu_i\to\bar\nu_j +X$ mode.
In the next section, we only consider the model, whose upper limit is
the most severe among all the models considered, i.e., the decay mode
$\bar\nu_i\to\bar\nu_j +X$.

In general, the decay redshifts depend on the neutrino energy since
the lifetime at the laboratory frame $\tau_{\rm lab}$ relates to that
at the neutrino rest frame $\tau$ via a simple relation $\tau_{\rm
lab}=E\tau/m$.
However, we believe that it does not make sense in discussing this
point strictly, because the estimation of the SRN flux contains many
other uncertainties as shown in Section \ref{sub:Supernova and supernova
rate model uncertainties}.
In order to obtain the lifetime of the mass eigenstates $\bar\nu_i$ from
the decay redshifts, the typical neutrino energy $E=10$ MeV is assumed
with the following formulation:
\begin{eqnarray}
 \tau_i&=&\frac{m}{E}\int_{z_i^d}^0\frac{dt}{dz}dz\nonumber\\
  &=&\frac{m}{E}\int_0^{z_i^d}\frac{dz}{H_0(1+z)}\nonumber\\
 &&{}\times\frac{1}{\sqrt{(1+\Omega_mz)(1+z)^2
  -\Omega_\Lambda (2z+z^2)}},
  \label{eq:lifetime}
\end{eqnarray}
where the Hubble constant $H_0$ is taken to be 70 km s$^{-1}$
Mpc$^{-1}$, and $\Lambda$-dominated cosmology is assumed ($\Omega_m=0.3,
\Omega_\Lambda =0.7$).
Since the exact value of the neutrino mass $m$ is not known, the
relevant quantities is the neutrino lifetime divided by its mass,
$\tau/m$.

\section{Results \label{sec:Results}}

\begin{figure}[htbp]
\begin{center}
\includegraphics[width=12cm]{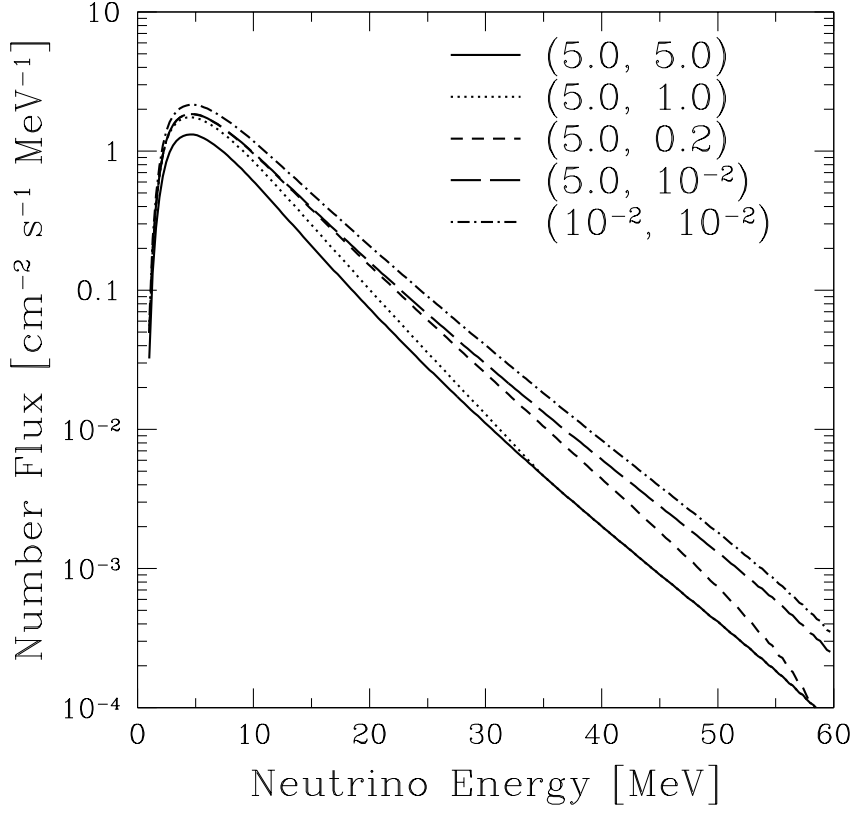}
\caption{The SRN flux for various parameter sets of decay
 redshifts. Each label represents
 $(z_2^d,z_3^d)$. \label{fig:srn}}
\end{center}
\end{figure}

Figure \ref{fig:srn} shows the SRN flux for various parameter sets of
decay redshifts $(z_2^d,z_3^d)$ as a function of neutrino
energy, which is calculated using Eqs. (\ref{eq:SRN flux}) and
(\ref{eq:decayed spectrum}).
The solid curve in Fig. \ref{fig:srn} shows the SRN flux in the case
that $(z_2^d,z_3^d)=(5.0,5.0)$, which represents the same flux as that
without the neutrino decay; this SRN flux almost coincides with one
obtained by the AST calculation.
Then, we included the decay of the heaviest mass eigenstate
$\bar\nu_3$ by reducing the value of $z_3^d$, with keeping
$\bar\nu_2$ stable.
When $z_3^d=1.0$ (dotted curve), the SRN flux at low energy region
$(E_{\bar\nu_e}\alt 35~{\rm MeV})$ deviates from the pure oscillation
model $(5.0,5.0)$.
This is because the neutrinos from supernovae at redshift larger than
$z_3^d=1.0$ are affected by the $\bar\nu_3\to\bar\nu_1$ decay and it
results in the increase of $\bar\nu_e$.
Since the neutrino energies are redshifted by a factor of $(1+z)^{-1}$
owing to an expansion of the universe, the decay effect can be seen at
low energy alone.
When the value of $z_3^d$ is reduced to $10^{-2}$, the neutrinos even
from the nearby sources are influenced by the $\bar\nu_3\to\bar\nu_1$
decay, resulting in the deviation over the entire energy range as shown
by the long-dashed curve in Fig. \ref{fig:srn}.
If we add the $\bar\nu_2\to\bar\nu_1$ decay, it further enhances the SRN
flux.

\begin{table*}[htbp]
\caption{The predicted SRN flux for various decay models and the
 corresponding SK limit (90\% C.L.). Integrated energy range is
 $E_{\bar\nu_e}>19.3$ MeV. The ratio of the prediction and the
 limit is shown in the fifth column. \label{table:limit}}
\begin{tabular}{ccccc}\hline
 Model $(z_2^d,z_3^d)$ & ($\tau_2/m,\tau_3/m$) [s/eV] & Predicted flux
 & SK limit (90\% C.L.) & Prediction/Limit\\ \hline
 $(5.0,5.0)$ & $(3.9\times 10^{10},3.9\times 10^{10})$ & 0.43 cm$^{-2}$
 s$^{-1}$ & $<1.2$ cm$^{-2}$ s$^{-1}$ & 0.35\\
 $(5.0,1.0)$ & $(3.9\times 10^{10},2.4\times 10^{10})$ & 0.55 cm$^{-2}$
 s$^{-1}$ & $<1.3$ cm$^{-2}$ s$^{-1}$ & 0.42\\
 $(5.0,0.2)$ & $(3.9\times 10^{10},7.7\times 10^9)$ & 0.93 cm$^{-2}$
 s$^{-1}$ & $<1.2$ cm$^{-2}$ s$^{-1}$ & 0.75\\
 $(5.0,10^{-2})$ & $(3.9\times 10^{10},4.4\times 10^8)$ & 1.0 cm$^{-2}$
 s$^{-1}$ & $<1.2$ cm$^{-2}$ s$^{-1}$ & 0.88\\
 $(10^{-2},10^{-2})$ & $(4.4\times 10^8,4.4\times 10^8)$ & 1.4 cm$^{-2}$
 s$^{-1}$ & $<1.2$ cm$^{-2}$ s$^{-1}$ & 1.2\\
 \hline
\end{tabular}
\end{table*}

In Table \ref{table:limit}, we summarize the SRN flux integrated over
the energy range of $E_{\bar\nu_e}>19.3$ MeV, for the each decay model.
In the second column we placed the lifetime-to-mass ratio which
corresponds to each decay redshift, which is obtained using
Eq. (\ref{eq:lifetime}).
The corresponding 90\% C.L. upper limit given by the SK observation at
$E_{\bar\nu_e}>19.3$ MeV \citep{Malek03} and prediction-to-limit ratio
are also shown in the fourth and fifth columns, respectively.
The predicted flux becomes larger as the decay is included, while the
observational upper limit remains unchanged.
At first sight, it is expected that the model $(10^{-2},10^{-2})$ is
already excluded by the current observational data.

\begin{figure}[htbp]
\begin{center}
\includegraphics[width=12cm]{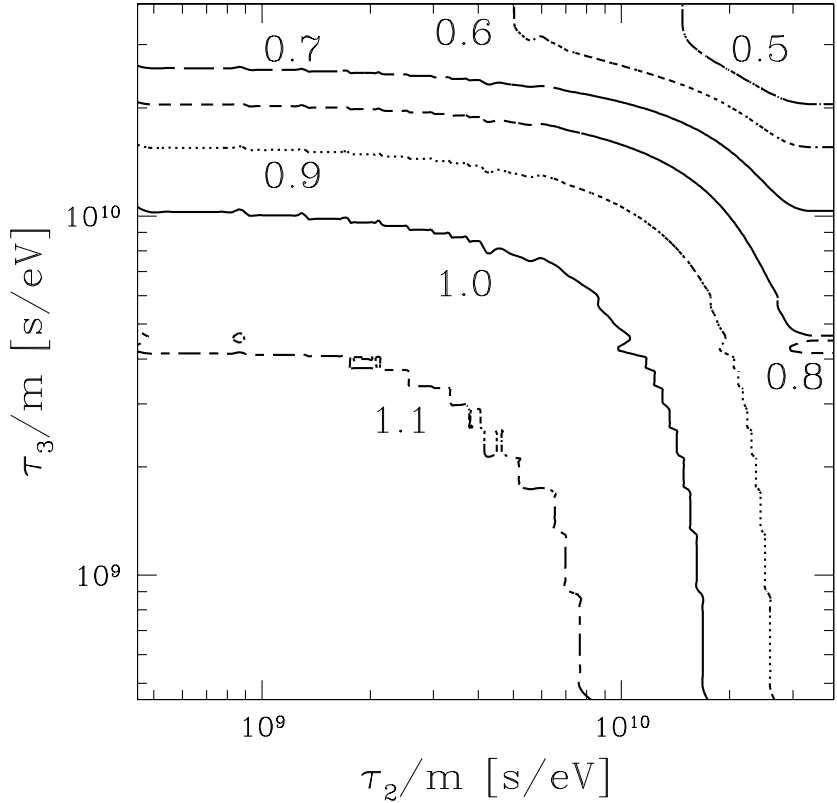}
\caption{A contour map of a prediction-to-limit ratio of the SRN
 flux which is projected against the lifetime-to-mass ratios
 $(\tau_2/m,\tau_3/m)$. \label{fig:ratio}}
\end{center}
\end{figure}

Figure \ref{fig:ratio} shows a contour plot for a ratio of the predicted
flux to the observational upper limit, which is projected against the
lifetime-to-mass ratios $(\tau_2/m,\tau_3/m)$.
The area below the solid curve labeled as 1.0 is considered to be an
excluded parameter region.
We can confirm that our approach is very powerful to obtain the
constraint on the neutrino lifetimes, because the best observed lower
limit thus far ($\agt 10^{-4}$ s/eV) is much smaller than that shown in
Fig. \ref{fig:ratio} ($\agt 10^{10}$ s/eV), although there are still
many uncertainties in this method as discussed in the next section.

\section{Discussion \label{sec:Discussion}}

\subsection{Other decaying modes \label{sub:Other decaying modes}}

Our discussions until this point were concerned with one specific
decaying mode, $\bar\nu_3\to\bar\nu_1,\bar\nu_2\to\bar\nu_1$, where the
daughter neutrinos carry nearly the full energy of their parent.
We consider the other possible decaying models; first $\bar\nu_3\to
\bar\nu_2,\bar\nu_2\to\bar\nu_1$ is discussed.
Since $\bar\nu_2$ state contains $\bar\nu_e$ state by less fraction than
$\bar\nu_1$, this decay mode gives smaller SRN flux, resulting in weaker
upper limit in general.
However, when the lifetime of one mode is much longer (shorter) than the
other, i.e., $z_2^d\ll z_3^d$ or $z_2^d\gg z_3^d$, the discussions for
our reference decay modes $\bar\nu_3\to\bar\nu_1,\bar\nu_2\to\bar\nu_1$
are basically applicable to this case.

When the daughter neutrino energy is considerably degraded, which is
actually the case when the neutrino masses are strongly hierarchical,
the observational upper limit for the each model is not as strong as the
previous limit shown in Table \ref{table:limit}.
This is because the energetically degraded daughter neutrinos soften the
SRN spectrum.

Finally we consider the case that the daughter neutrinos are sterile
species which does not interact with matter.
If the lifetime of the mode is sufficiently short, then the obtained
spectrum from each supernova can be expressed by
\begin{equation}
 \mathcal N_{\bar\nu_e}(E,z)\simeq
  \cos^2\theta_\odot\mathcal N_{\bar\nu_e}^0(E,z),
  \label{eq:sterile}
\end{equation}
which is smaller than the normal oscillation expression (\ref{eq:pure
oscillation}).
Thus, also in this case, the observational upper limit will be looser.
In consequence, all the other possible decaying models give weaker upper
limit compared to our reference model.

\subsection{Supernova and supernova rate model uncertainties
\label{sub:Supernova and supernova rate model uncertainties}}

Again we restrict our discussion to our standard decay mode, and discuss
whether the parameter region $(\tau_2/m,\tau_3/m)\alt (10^{10},10^{10})$
[s/eV] is really ruled out by the current observational data, as shown
in Fig. \ref{fig:ratio}.

The theoretical calculation of the SRN flux contains many uncertainties
such as the supernova rate as a function of redshift $z$ as well as the
original neutrino spectrum emitted by each supernova.
As for the supernova rate, since we have inferred it from the rest-frame
UV observation, it may be severely affected by dust extinction, which is
quite difficult to estimate.
Thus for example, if we use the supernova rate model which is larger by
a factor of 2 (even if this is actually the case, it is not surprising
at all), then almost all the region of lifetime-to-mass ratio
$(\tau_2/m,\tau_3/m)$ will be excluded as shown in
Fig. \ref{fig:ratio}.

The uncertainty concerning the original neutrino spectrum also gives
very large model dependence of the SRN flux calculation.
Although we adopted the result of the Livermore simulation, as we have
noted above it lacks the relevant neutrino processes, which reduce the
average energy of $\nu_x$ to the value close to that of $\bar\nu_e$.
If this is the case, the SRN spectrum as a result of the neutrino decay
becomes softer compared to that obtained with the Livermore original
spectra, which leads to weaker upper limit.
In actual, we have calculated the SRN flux for various values of
$(z_2^d,z_3^d)$ assuming the original $\nu_x$ spectrum is the same as
that of $\bar\nu_e$, for which we have used the Livermore data.
As a result of the calculation for this extreme case, the obtained
prediction-to-limit ratio is at most $\sim 0.35$, even when both of
decay redshifts are sufficiently small.
In consequence, considering the uncertainties which is included in the
supernova rate or the neutrino spectrum models, we cannot conclude that
the parameter region $(\tau_2/m,\tau_3/m)\alt (10^{10},10^{10})$ [s/eV]
is already excluded, although Fig. \ref{fig:ratio} indicates it.

\subsection{Future prospects \label{sub:Future prospects}}

Now, we consider the future possibility to set severer constraint on the
neutrino decay models.
The largest background against the SRN detection at SK is so-called
invisible muon decay products.
This event is illustrated as follows.
The atmospheric neutrinos produce muons by interaction with the nucleons
(both free and bound) in the fiducial volume.
If these muons are produced with energies below \v{C}herenkov radiation
threshold (kinetic energy less than 53 MeV), then they will not be
detected (``invisible muons''), but their decay-produced electrons will
be.
Since the muon decay signal will mimic the $\bar\nu_ep\to e^+n$
processes in SK, it is difficult to distinguish SRN from these events.
Recent SK limits are obtained by the analysis including this invisible
muon background.

In the near future, however, it should be plausible to distinguish the
invisible muon signals from the SRN signals; two different methods are
currently proposed for the SK detector.
One is to use the gamma rays emitted from nuclei which interacted with
atmospheric neutrinos \citep{Suzuki03}.
If gamma ray events, whose energies are about 5--10 MeV, can be
detected before invisible muon events by muon lifetime, we can subtract
them from the candidates of SRN signals.
In that case, the upper limit would be much lower (by factor $\sim 3$)
when the current data of 1,496 days are reanalyzed \citep{Suzuki03}.
Another proposal is to detect signals of neutrons, which are produced
through the $\bar\nu_ep\to e^+n$ reaction, in the SK detector.
Actual candidate for tagging neutrons is gadolinium solved into pure
water \citep{Vagins03}.
Additional neutron signals can be used as delayed coincidence to
considerably reduce background events.
In addition, future projects such as the Hyper-Kamiokande detector is
expected to greatly improve our knowledge of the SRN spectral shape as
well as its flux.
If a number of data were actually acquired, the most general and
model-independent discussions concerning the neutrino decay would be
accessible.
Finally,  we again stress that a great advantage to use the SRN is that
the cosmologically long lifetime can be probed {\it in principle}.

\section{Conclusions}

We obtained the current lower limit to the neutrino lifetime-to-mass
ratio using the SRN observation at SK.
Since the available data are only the upper limit on the flux integrated
over $E_{\bar\nu_e}>19.3$ MeV, the decay model on which we can set
constraints is quite restricted, i.e., the decay models that gives the
SRN flux which is comparable to or larger than the corresponding upper
limit.

Therefore, our reference decay model in this Letter is the two-body
decay $\bar\nu_i\to\bar\nu_j +X$, with the daughter neutrinos which are
active species and carry nearly full energy of their parent.
The neutrino mass hierarchy is assumed to be normal ($m_1<m_3$).
The SRN calculation is also very sensitive to the adopted supernova
model, i.e., neutrino spectrum from each supernova explosion; we adopted
the result of the numerical simulations of the Lawrence Livermore group.
Although their calculation is recently criticized since it lacks several
relevant neutrino processes, we adopt it as a reference model.
This is because it is only the successful model of the supernova
explosion and we definitely need fully time-integrated neutrino spectra.

Our calculations with these models shows that the neutrino decay model
with $(\tau_2/m,\tau_3/m)\alt (10^{10},10^{10})$ [s/eV] apparently gives
the SRN flux that is larger than the current upper limit (see
Fig. \ref{fig:ratio}).
Since this value $10^{10}$ s/eV is much larger than that of the limit
obtained by the solar neutrino observation $\agt 10^{-4}$ s/eV, our
approach is shown to be very powerful to obtain the implications for the
neutrino decay.
At present, however, owing to the large uncertainties such as supernova
models, this lower limit $10^{10}$ s/eV cannot be accepted without any
doubt.
Future experiments with the updated detectors and reanalysis with the
refined method is expected to give us greatly improved information on
the SRN flux and spectrum.
In that case, the most general model-independent discussions concerning
the neutrino decay would be possible.

\begin{acknowledgments}
The author is supported by Grant-in-Aid for JSPS Fellows.
\end{acknowledgments}

\bibliography{refs}

\end{document}